\documentclass[twocolumn]{aastex631}
\usepackage{amsmath}
\usepackage[mathlines]{lineno}

\begin{document}

\title{Dynamics of Star Cluster Formation: The Effects of Ongoing Star Formation and Stellar Feedback}

\correspondingauthor{Jeremy Karam}
\email{karamj2@mcmaster.ca}

\author[0000-0003-3328-329X]{Jeremy Karam}
\affiliation{Department of Physics and Astronomy, McMaster University, 1280 Main Street West, Hamilton, ON, L8S 4M1, Canada}

\author[0000-0002-6465-2978]{Michiko S. Fujii}
\affiliation{Department of Astronomy, Graduate School of Science, The University of Tokyo, 7-3-1
Hongo, Bunkyo-ku, Tokyo 113-0033, Japan}

\author[0000-0003-3551-5090]{Alison Sills}
\affiliation{Department of Physics and Astronomy, McMaster University, 1280 Main Street West, Hamilton, ON, L8S 4M1, Canada}

%% Note that the \and command from previous versions of AASTeX is now
%% depreciated in this version as it is no longer necessary. AASTeX 
%% automatically takes care of all commas and "and"s between authors names.

%% AASTeX 6.31 has the new \collaboration and \nocollaboration commands to
%% provide the collaboration status of a group of authors. These commands 
%% can be used either before or after the list of corresponding authors. The
%% argument for \collaboration is the collaboration identifier. Authors are
%% encouraged to surround collaboration identifiers with ()s. The 
%% \nocollaboration command takes no argument and exists to indicate that
%% the nearby authors are not part of surrounding collaborations.

%% Mark off the abstract in the ``abstract'' environment. 
\begin{abstract}

We perform a high resolution zoom-in simulation of star cluster assembly including the merger of two sub-clusters with initial conditions taken from previous large scale giant molecular cloud (GMC) simulations. We couple hydrodynamics to N-body dynamics to simulate the individual stars themselves, and the gas-rich environment in which they evolve. We include prescriptions for star formation and stellar feedback and compare directly to previous simulations of the same region without these prescriptions to determine their role in shaping the dynamics inherited from the cluster assembly process. The stellar mass of the cluster grows through star formation within the cluster and accretion of new stars and star forming gas from a nearby filament. This growth results in an enhancement in the cluster's rotation and anisotropic expansion compared to simulations without star formation. We also analyze the internal kinematics of the cluster once it has lost most of its gas and find that the rotational velocity and the velocity anisotropy profiles are qualitatively similar to those expected of clusters that have undergone violent relaxation. As well, rotation and anisotropic expansion are still present by the time of gas removal. This implies that evolution within the GMC was unable to completely erase the kinematics inherited by the merger.

\end{abstract}

%% Keywords should appear after the \end{abstract} command. 
%% The AAS Journals now uses Unified Astronomy Thesaurus concepts:
%% https://astrothesaurus.org
%% You will be asked to selected these concepts during the submission process
%% but this old "keyword" functionality is maintained in case authors want
%% to include these concepts in their preprints.

%% From the front matter, we move on to the body of the paper.
%% Sections are demarcated by \section and \subsection, respectively.
%% Observe the use of the LaTeX \label
%% command after the \subsection to give a symbolic KEY to the
%% subsection for cross-referencing in a \ref command.
%% You can use LaTeX's \ref and \label commands to keep track of
%% cross-references to sections, equations, tables, and figures.
%% That way, if you change the order of any elements, LaTeX will
%% automatically renumber them.
%%
%% We recommend that authors also use the natbib \citep
%% and \citet commands to identify citations.  The citations are
%% tied to the reference list via symbolic KEYs. The KEY corresponds
%% to the KEY in the \bibitem in the reference list below. 

\section{Introduction} \label{sec:intro}

Simulations show that star cluster formation is a hierarchical process involving the mergers of smaller star clusters into larger ones while embedded inside giant molecular clouds (GMCs) (e.g. \citealt{Howard2018}, \citealt{chen}, \citealt{rieder21}). Observationally, one cannot pierce through the extinction caused by GMC gas to see the stars themselves, so evidence for hierarchical star cluster buildup comes from dynamical studies of star clusters that have recently removed their surrounding gas through a combination of stellar feedback and star formation. For example, observations of cluster age gradients (e.g. \citealt{fahrion}), asymmetric dynamics (e.g. \citealt{wright}, \citealt{wright_2019}, \citealt{wright_2}, \citealt{armstrong_tan}), and preferential directions of high velocity (runaway) stars (e.g. \citealt{Polak2024}, \citealt{stoop_r136}) are interpreted as dynamical signatures of mergers as a build up mechanism for star clusters (e.g. \citealt{schoettler_22}, \citealt{lahen_24}, \citealt{Karam2024}, \citealt{CCC_JK}). To completely draw this connection between star cluster formation inside GMCs and the observations of star clusters after they have left their natal cloud, it is important to clearly understand all of the complexities associated with star cluster build up and the environment within which it takes place.

As star clusters merge inside a GMC, gas around them becomes compressed and can go on to form new stars (\citealt{karam}, \citealt{fujii_sirius_4}). New star formation, coupled with the increase in mass from the merger of two clusters can increase the mass of clusters involved in the merger. However, mergers can also result in the ejection of stars from their host cluster (\citealt{karam}) and cluster splitting (\citealt{dobbs_22}) which both lead to a decrease in the total cluster mass implying a complex change to the cluster dynamics throughout its assembly. The formation of new stars from cluster mergers may also offer an explanation for the age gradients observed in many clusters and associations, but the origin of the kinematic substructure among these groups (e.g. \citealt{gaudin_puppis}, \citealt{zari_orion}, \citealt{US_kin_subgroups}) is still unclear. It has been shown that kinematic distinction may arise from new stars inheriting motions from dense gas (\citealt{ngc_6357}) but it is uncertain whether this distinction remains after the surrounding gas distribution has been altered by stellar feedback. Regarding the role played by stellar feedback, \citet{lewis_FB} showed that feedback from massive stars can not only alter the amount of star formation present throughout a GMC, but also the spatial distribution of resultant clusters. Similarly, \citet{dobbs_1} found that the rate of hierarchical cluster buildup can change depending on environment inside a star forming spiral galaxy implying that feedback may be able to alter the efficiency of star cluster assembly inside a GMC. As pre-SNe feedback has been shown to be dominant in removing gas from GMCs and altering the environment (e.g. \citealt{clearing_time}), it is likely to have an effect on the dynamics of clusters as star formation is ongoing.

A key component to the morphology of the gaseous environment of star forming regions is dense filaments (see \citealt{hacar_fils} for a review). These filaments are found around young clusters in the form of hub filament systems (e.g. \citealt{wong_22}, \citealt{kumar}, \citealt{dewangan}, \citealt{seshadri}, \citealt{rawat}, \citealt{zhang_hfs}, \citealt{zhang_w49}) which can be responsible for the host cluster's increase in mass and size through accretion onto the cluster (\citealt{kirk}, \citealt{karam_23}). As well, recent observation from \citet{hfs_g321} suggests convergence of filamentary flows onto dense, massive star forming clumps implying that further evolution of hub filament systems may lead to the formation of star clusters. Understanding the flow of gas onto clusters is necessary to understand the dynamics of the cluster itself because of the star formation that will take place along the filamentary flows.

In \citet{Karam2024} (hereafter Paper I), we showed that the surrounding gas environment plays a vital role in the build up of clusters through facilitating mergers at early times in the GMCs life. It is therefore important to consider realistic gas environments alongside stellar feedback and star formation to gain a full picture of the impacts of star cluster mergers on the dynamics of the clusters involved. In this work, we take a previous simulation performed in Paper I, and include prescriptions for star formation and stellar feedback to concretely discern their influence on the merger product. We analyze how both processes affect the phase space of the stars belonging to the cluster during and after the merger process. We also analyze the kinematics of the cluster after it has removed most of its surrounding gas. This paper is structured as followed: in Section \ref{sec:methods} we discuss the numerical methods and initial conditions used in our simulation, in Section \ref{sec:results} we discuss the evolution of the stars and gas in our star cluster merger simulation, in Section \ref{sec:discussion}, we discuss the implications of our results, and in Section \ref{sec:summary}, we summarize our key results.

\begin{figure*}
    \centering
    \includegraphics[scale=0.5]{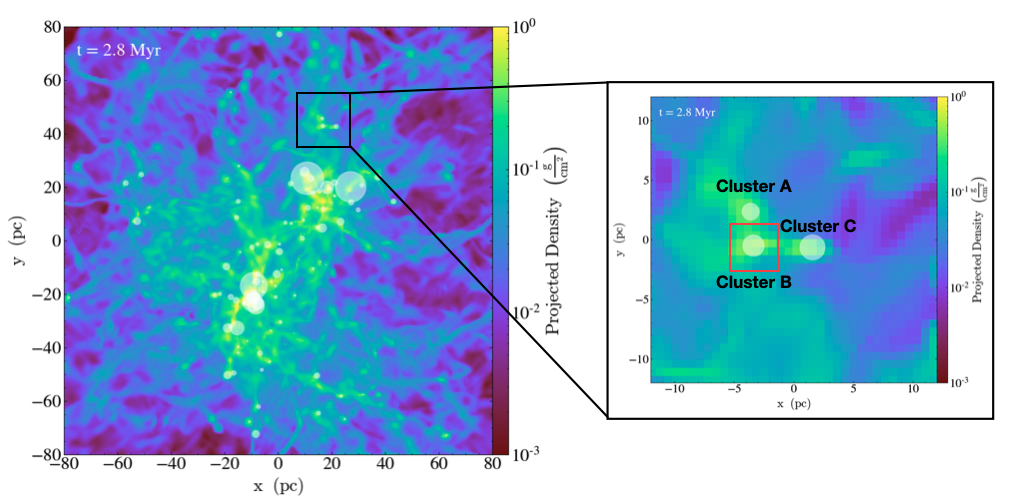}
    \caption{Left: Snapshot from simulation performed in \citet{Howard2018}. White circles show locations of the sink particles with their size representing the mass of the sink. Black box shows the region taken for simulation \texttt{region1} in Paper I and \texttt{region1\_SFFB} in this work. Colour shows gas surface density. Right: Zoom-in of region used as initial conditions for \texttt{region1} simulation (see Figure \ref{fig:pynbody_spur}). Red box shows the sink particle whose stellar mass is plotted as the red line in Figure \ref{fig:bound_mass}.}
    \label{fig:fil_IC}
\end{figure*}

\section{Methods}\label{sec:methods}
\subsection{Numerical Methods}
\label{sec:numerical_methods}

We perform our simulations using the ASURA+BRIDGE code (\citealt{fujii_sirius_2}, \citealt{fujii_sirius_3}). ASURA+BRIDGE evolves gas using the smoothed particle hydrodynamics (SPH) code \texttt{ASURA} (\citealt{saitoh_2008}, \citealt{saitoh_2009}), N-body dynamics using \texttt{PETAR} (\citealt{Wang2020b}) and connects the two using the \texttt{BRIDGE} scheme (\citealt{bridge}). The two codes are connected every bridge timestep which we choose to be $\Delta t_B$ = 200yr. In \texttt{PETAR}, forces from distant particles are calculated using a tree method (\citealt{tree}) and nearby particle forces are calculated using a Hermite scheme. As well, \texttt{PETAR} includes a slow down algorithmic regularization scheme allowing the evolution of multiple systems and close encounters (\citealt{Wang2020a}).

ASURA+BRIDGE also includes a star formation prescription outlined in \citet{fujii_sirius_1} and a feedback model outlined in \citet{fujii_sirius_3}. Stars form throughout our simulations with a given efficiency per free fall time. Once gas is above a certain density threshold, below a certain temperature threshold, and is converging, it is eligible to form a star. We choose the efficiency per free fall time to be 0.02, the density threshold to be 10$^{4}$cm$^{-3}$, and the temperature threshold to be 30K. Once the above conditions are satisfied, a stellar mass is randomly chosen from a \citet{kroupa2001} mass function between 0.1M$_\odot$ and 150M$_\odot$. To form the star, we assemble mass from the surrounding region within a given radius. We choose this radius to be 0.2pc because \citet{fujii_sirius_1} shows that this value results in star formation that fully samples the \citet{kroupa2001} mass function. The position and velocity of the newly formed star are inherited from the gas used to create the star.

Feedback is modelled through HII regions around massive stars. Once stars are massive enough, a Str\"{o}mgren (\citealt{stromgren}) radius is estimated around the star using local density and ionizing photon rates from \citet{lanz_hubeny}. Gas inside the radius is injected with mechanical energy in the form of momentum to account for radiation (\citealt{kim_rad}) and stellar winds (\citealt{renaud_wind}). As well, thermal energy is injected into the surrounding gas to raise the temperature to 10$^4$K. Such feedback is implemented for stars more massive than 10M$_\odot$.

\subsection{Initial Conditions}
We take our initial conditions from previously run GMC simulations performed in \citet{Howard2018} (hereafter H18) using the \texttt{Flash} grid code (\citealt{Fryxell2000}) and convert the background gas into an SPH representation using the method outlined in Paper I. In summary, we take each grid cell present in the region of interest in the H18 simulation and place N$_{\mathrm{SPH}}$ = M$_{\mathrm{cell}}$/m$_{\mathrm{SPH}}$ (where M$_{\mathrm{cell}}$ is the mass of the given grid cell, and m$_{\mathrm{SPH}}$ is the initial SPH particle mass) gas particles in a Gaussian distribution around the centre of each grid cell. The velocity of the gas in each grid cell is assigned to every SPH particle placed in that grid cell. We also give all gas in the simulation a velocity dispersion of 10kms$^{-1}$ which is consistent with simulations of star forming regions (\citealt{pillsworth}). Lastly, we provide every SPH particle with the same internal energy as that of the grid cell it is placed around.

In the region chosen for this work, there are three sink particles, two of which eventually merge. Each sink contains a stellar and a gas component. We convert the sink particles to star clusters in a similar way as that presented in \citet{karam}. We initialize the stellar component as a Plummer (\citealt{plummer1911}) sphere. We choose the size of the stellar distribution such that the density at the half mass radius is consistent with young massive cluster observations (\citealt{zwart}). We choose the velocities of the stars such that the cluster is initially in virial equilibrium. We let the stellar distribution relax in a background gas distribution whose mass is given by the sink particle gas mass before placing the stars in the larger simulation alongside the background gas. We initialize the gas component of each sink as a uniform distribution in the larger simulation. We choose the size of the gas distribution such that the average density of the sink particle gas is 10$^{4}$cm$^{-3}$ which is the threshold value for sink formation in H18. We give the gas a velocity dispersion of 10kms$^{-1}$ to match the background gas. The region chosen is the same as \texttt{region1} from \citet{Karam2024} allowing us to directly test the roles played by feedback and star formation (see Figure \ref{fig:fil_IC}). This region contains three clusters (cluster A, B, and C). Clusters A and B participate in the merger, while cluster C does not. We call the simulation in this work \texttt{region1\_SFFB}.

\begin{figure*}
    \centering
    \includegraphics[scale=0.23]{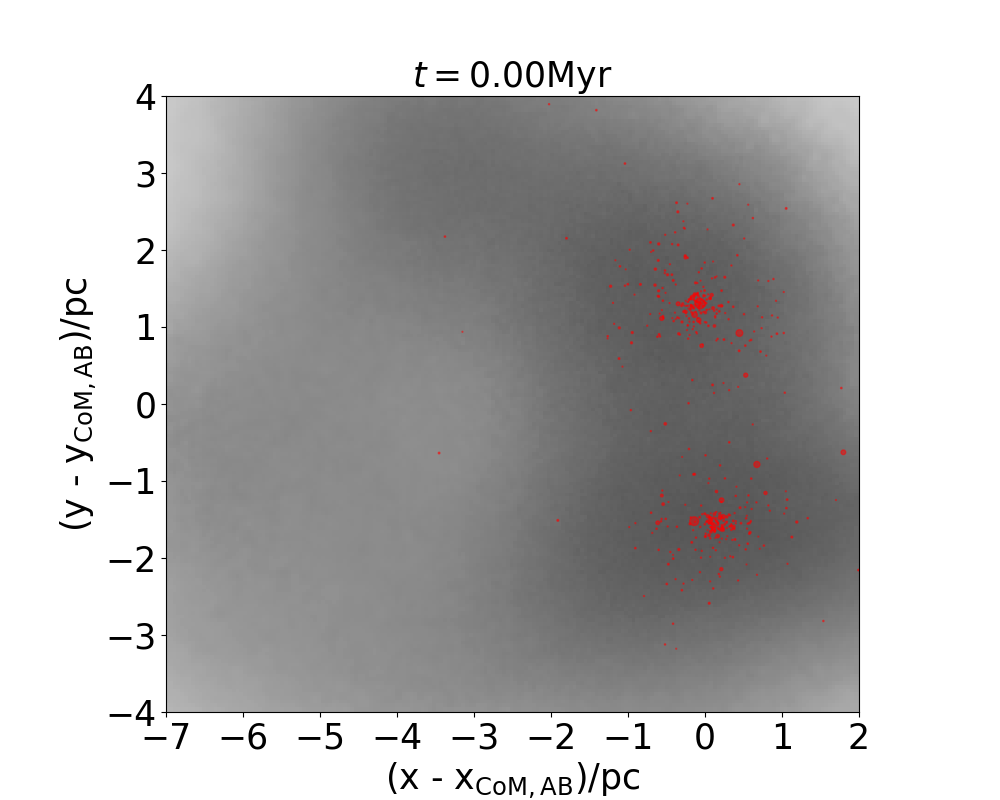}
    \includegraphics[scale=0.23]{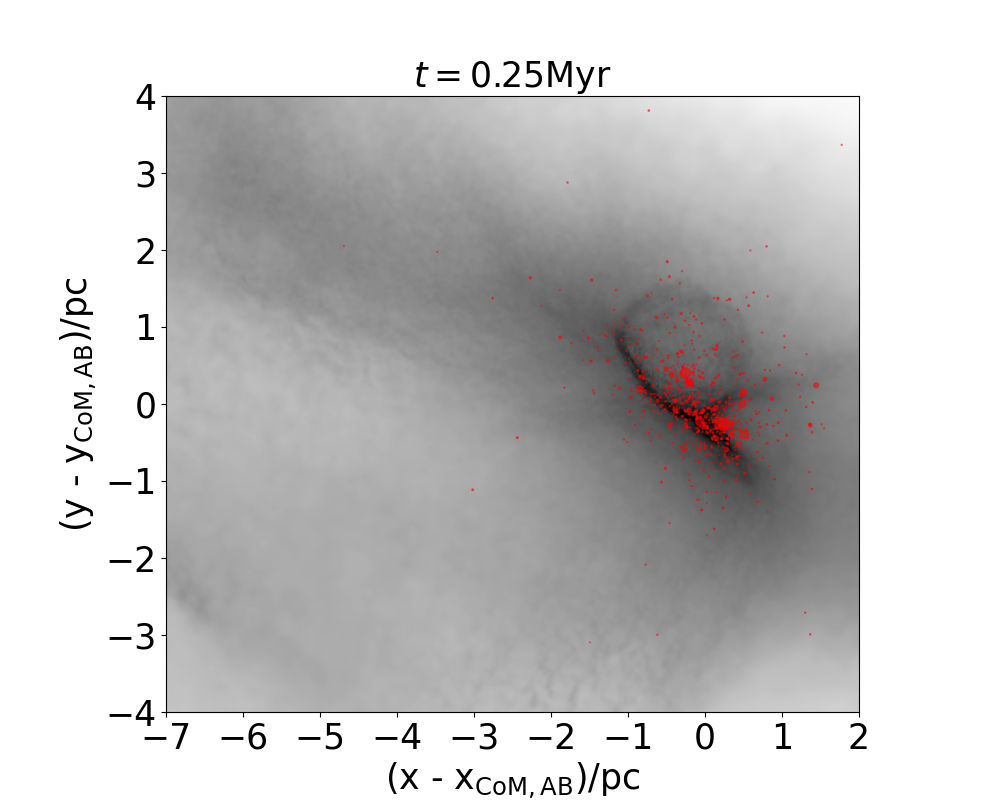}
    \includegraphics[scale=0.23]{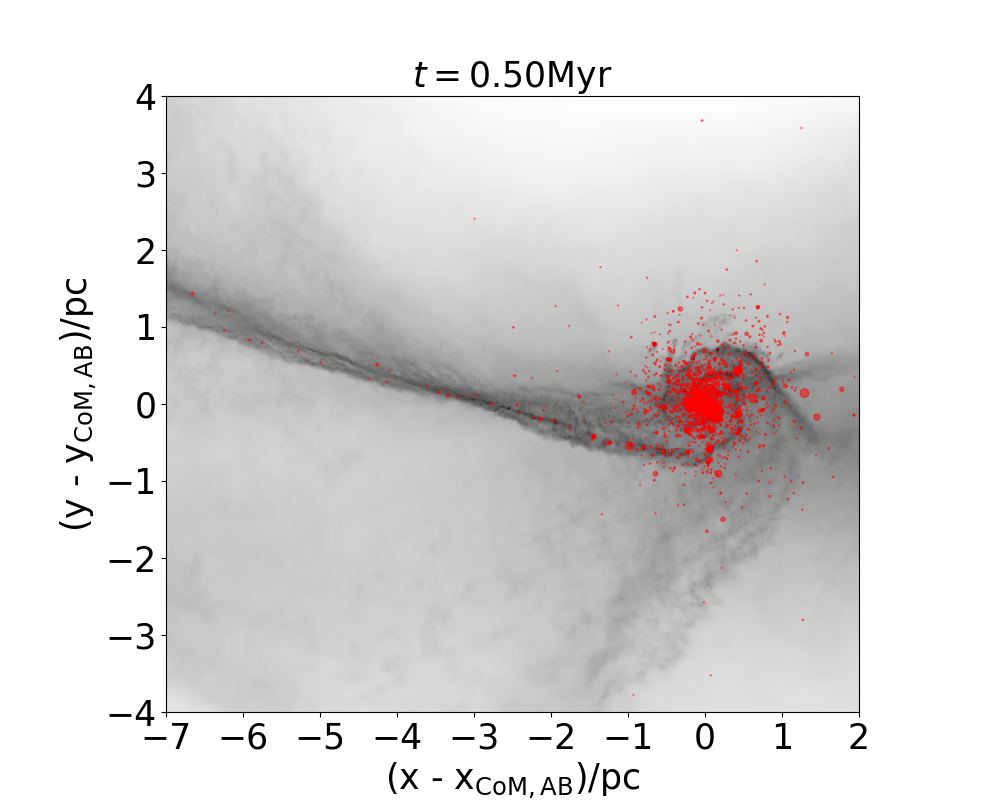}
    \includegraphics[scale=0.23]{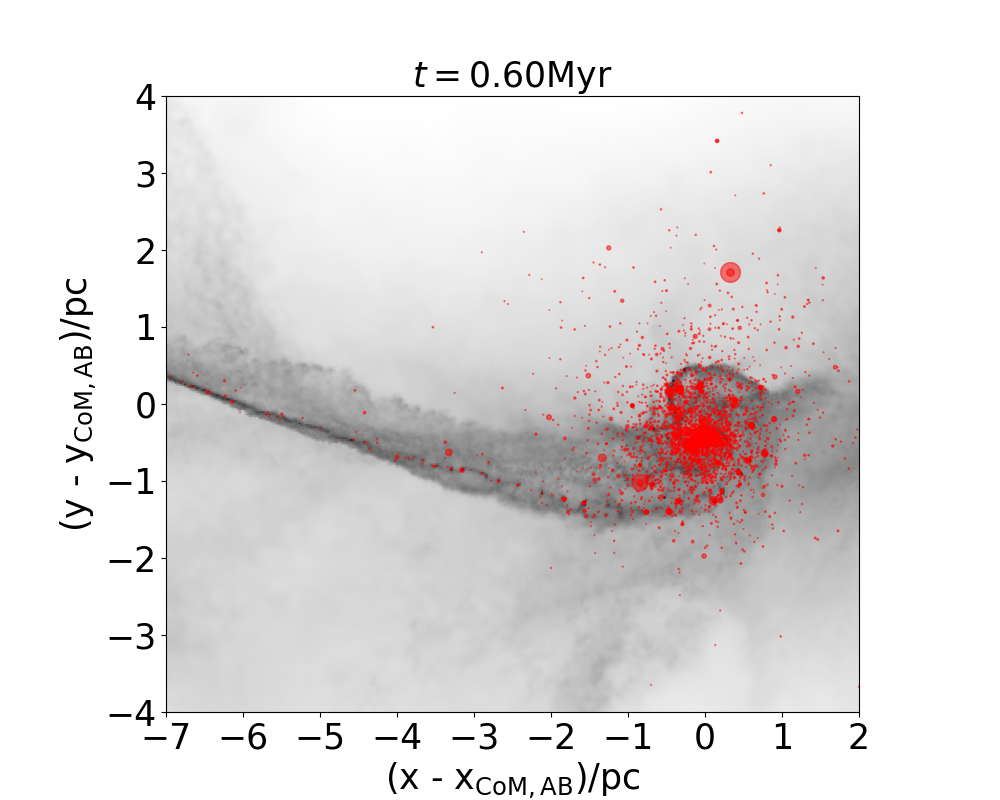}
    \includegraphics[scale=0.23]{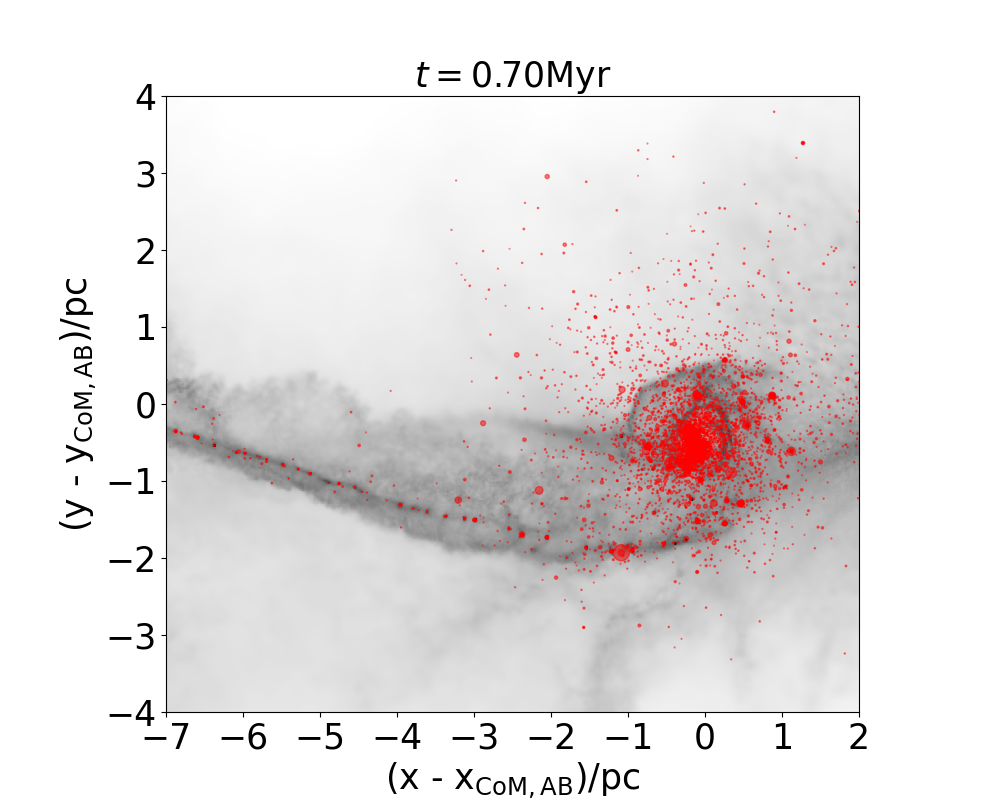}
    \includegraphics[scale=0.23]{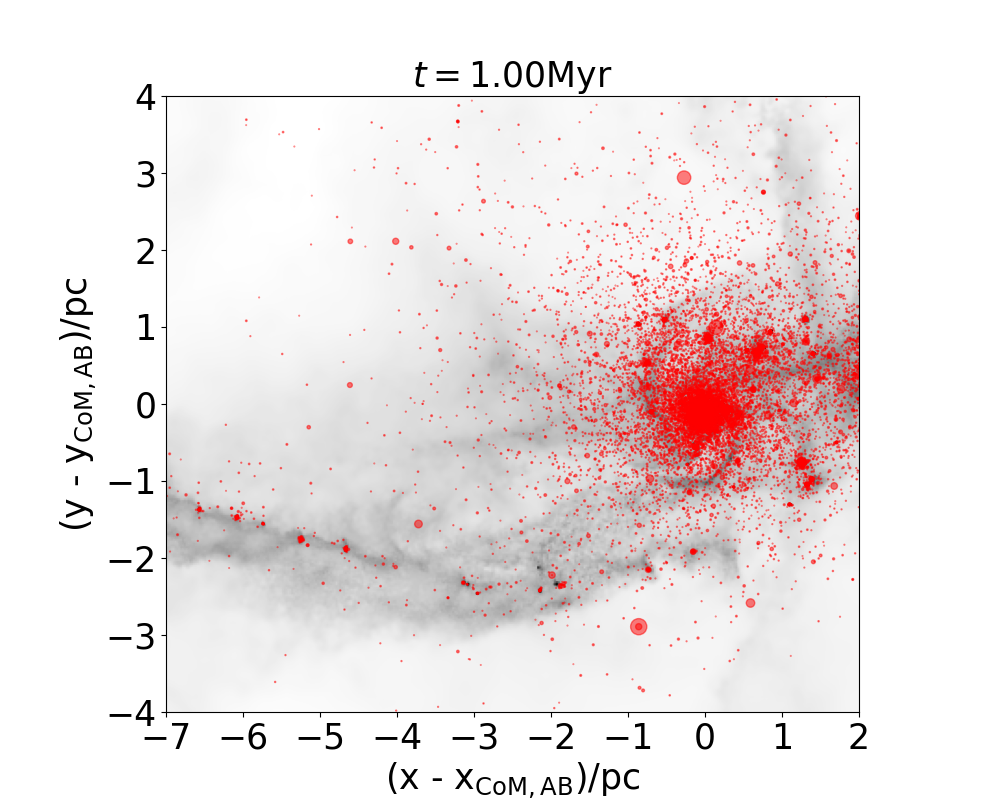}
    \caption{Snapshots of gas and bound stars around cluster AB from the \texttt{region1\_SFFB} simulation. The x and y axes show the x and y positions subtracted by the centre of mass of cluster AB. Red circles represent the stars in each cluster and their size scales with the stars mass. Gas is shown in black with darker regions showing gas with higher densities. The range in gas densities shown is 0.1$-$10$^3 $M$_\odot$pc$^{-3}$.}
    \label{fig:pynbody_spur}
\end{figure*}

\subsection{Identifying Star Clusters}
Throughout the simulations, as clusters merge, and new stars form, the membership of our star clusters changes. We keep track of the components of each star cluster using a similar method as that outlined in \citet{Karam2024}. 

We first use HDBSCAN (\citealt{hdbscan}) to cluster the stars in 3D position space. To be considered part of a given cluster, we ensure that the spatially clustered stars are gravitationally bound. We then check whether the spatially unclustered stars are bound to each of the clusters present. If they are, we assign those spatially unclustered stars to the cluster with which it is most bound. If they are not bound to any cluster, we consider that star unbound. When checking for boundedness, we also consider all of the gas present in the simulation.

\begin{figure*}
    \centering
    \includegraphics[scale=0.43]{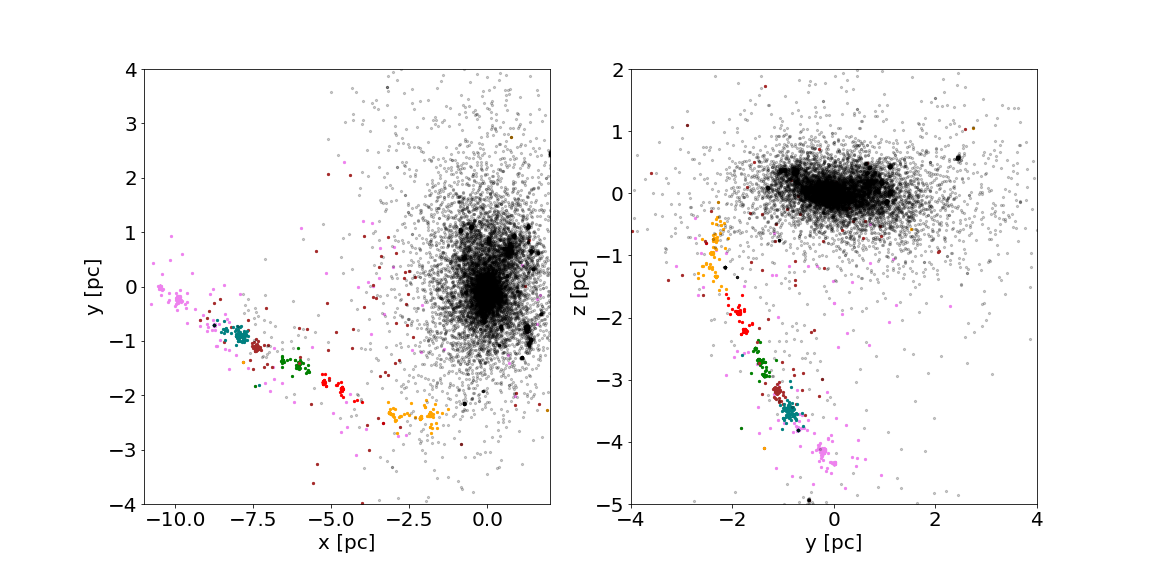}
    \caption{The locations of stars at t$=$1Myr after the beginning of the \texttt{region1\_SFFB} simulation. The colours of the points indicate which subgroup each star belongs to as determined by \texttt{HDBSCAN}. The black points are made more translucent than the coloured points to make it easier to see the coloured clumps.}
    \label{fig:hdbscan}
\end{figure*}

\section{Results}
\label{sec:results}

\begin{figure}
    \centering
    \includegraphics[width=1\linewidth, height=0.95\linewidth]{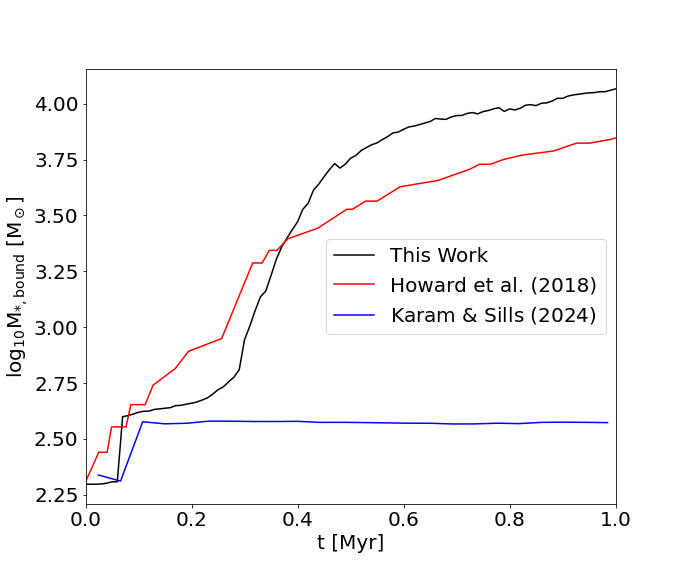}
    \caption{Bound stellar mass of cluster AB in \texttt{region1\_SFFB} from this paper (black) and \texttt{region1} from Paper I (blue). Red line shows the mass of the most massive sink that participates in the merger from the region in the H18 simulation (see Figure \ref{fig:fil_IC}).}
    \label{fig:bound_mass}
\end{figure}

We show the evolution of the \texttt{region1\_SFFB} simulation in Figure \ref{fig:pynbody_spur}. In this simulation, cluster A begins merging with cluster B at t$\approx0.06$Myr which is roughly the same as the merger time in the \texttt{region1} simulation from Paper I (see Figure \ref{fig:bound_mass}). We call the merged cluster ``cluster AB".

To discern the role played by the newly added physics (star formation, feedback, and cooling), we analyze cluster AB at t$=$1Myr and compare it to cluster AB at the same time from the \texttt{region1} simulation from Paper I. We show the distribution of bound stars at t$=$1Myr in the \texttt{region1\_SFFB} simulation in Figure \ref{fig:hdbscan}. Stars belonging to cluster AB are shown as black and grey points while coloured clumps can be seen along the negative x and z axes. These coloured clumps are the result of star formation taking place along a filament that is feeding gas to  cluster AB throughout the simulation (this filament can be seen in the top-right panel of Figure \ref{fig:pynbody_spur}, and we discuss it more in Section \ref{sec:fils}). The substructure we see in position space from Figure \ref{fig:hdbscan} is not present in velocity space. All of the stars belonging to cluster AB are mixed in velocity space as expected from a merger (\citealt{Karam2024}). As well, the clumps which form along the filament are mixed with each other and with cluster AB in velocity space at t$=$1Myr.

Figure \ref{fig:bound_mass} shows the mass of cluster AB as a function of time after the beginning of the simulation. We see here that the bound stellar mass of the cluster increases drastically as the merger takes place thanks to star formation. We compare this change in mass to that of the original sink particles from the H18 simulation as shown by the red line in Figure \ref{fig:bound_mass} which shows the bound stellar mass of the most massive sink that participates in the merger as a function of time (see Figure \ref{fig:fil_IC}). The merger in the H18 simulation happens at a later time than in this work. In the H18 simulations, a sink particle converted gas into stars at a constant rate of 20\% every free-fall time (0.36Myr). Our simulations show that the rate at which cluster AB gains bound stellar mass is not constant. This, along with our simulations ability to resolve gas at higher resolution than H18, results in our cluster having an overall higher mass than the corresponding H18 sink by t$=$1Myr. The mass of the H18 sink is $\approx$ 78\% of the mass of cluster AB at t$=$1Myr. Lastly, the mass of cluster AB in the \texttt{region1} simulation stays constant after the merger takes place (blue line in Figure \ref{fig:bound_mass}). This is because we did not include star formation in this simulation from Paper I.

There are two main processes contributing to this increase in mass in our simulation. The first is star formation taking place within cluster AB from the newly compressed gas. The second is the accretion of stellar clumps and star forming gas from the filament that traveled into cluster AB. Both of these processes together result in an increase in the bound stellar mass by $\approx$2.5 orders of magnitude in the \texttt{region1\_SFFB} simulation when compared to the \texttt{region1} simulation in Paper I.

\subsection{Ejected Stars From In-Situ Star Formation}
With ASURA+BRIDGE, we resolve star formation and the formation of dynamical binaries but do not account for primordial binaries. The simulation begins with 31 binary systems in total. There are 9 and 7 binary systems in cluster A and B respectively at the beginning of the simulation. In the \texttt{region1\_SFFB} simulation, dynamical binaries form as a result of the high density environment created by the merger of cluster A with cluster B. 

Before the merger takes place, the binary fraction stays roughly constant. Consequently, only 1 star becomes unbound before the merger begins. Throughout the simulation after the merger, the number of binary systems increases. By t$=$1Myr, 5\% of all stars belonging to cluster AB are in a binary system. Out of all the binary systems present in the simulation, 71\% of them are in cluster AB at t$=$1Myr. The high concentration of binaries in cluster AB results in the dynamical ejection of stars and increases the unbound fraction of stars after the merger takes place. At t$=$1Myr, $\approx$ 25\% of the stars that formed in cluster AB have become unbound. This is higher than the unbound fraction found as a result of mergers in both \citet{karam}, and \citet{CCC_JK} implying that, while the merger process itself can increase the fraction of unbound stars around a cluster, new star formation in the dense environment created by the cluster merger results in a higher unbound fraction through dynamical ejections.

\begin{figure}
    \centering
    \includegraphics[scale=0.4]{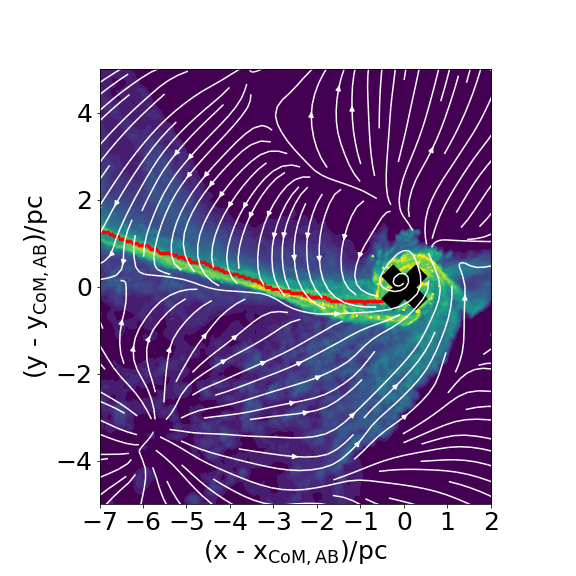}
    \caption{Dynamics of the gas around cluster AB in the \texttt{region1\_SFFB} simulation at t$=$0.5Myr. Red line shows the identified filament using \texttt{fil\_finder}, and the white stream lines show the density-weighted velocities of the gas. The gas density is shown as the colourmap with yellow regions showing denser gas. The range of gas densitiies is 10-10$^5$M$_\odot$pc$^{-3}$. The black cross shows the location of the centre of mass of cluster AB.}
    \label{fig:vel_fil}
\end{figure}

\subsection{Filament Accretion}
\label{sec:fils}
We use the \texttt{python} packages \texttt{fil\_finder} (\citealt{fil_finder}) and \texttt{radfil} (\citealt{radfil}) to identify filamentary shapes and calculate filament parameters in our simulations. To employ \texttt{fil\_finder}, we convert our SPH particle data into a grid using \texttt{pynbody} (\citealt{pynbody}) to average the gas density 5pc along the z-axis around cluster AB to ensure we are not including gas that belongs to cluster C. 

We find that a filament is well defined at t$\approx$0.5Myr as seen in Figure \ref{fig:vel_fil}. The width of the filament as given by the width of the Gaussian best fit is $\approx 0.4$pc at this time in the simulation. We also show the dynamics of the gas as streamlines in the same figure and find that after the start of the merger, gas along the filament has net motions towards cluster AB. Furthermore, the merger has offset the densest component of the filament from the centre of mass of cluster AB and has wrapped around the cluster which imparts a large amount of angular momentum onto the stars in cluster AB (we discuss this more in Section \ref{sec:dynamics_r1}). As most of the stellar mass belonging to cluster AB by t$=$1Myr comes from newly formed stars, and these newly formed stars inherit the motions of the gas they form from, the dynamics of the gas has a direct impact on the dynamics of stellar component of the cluster.

At t=1Myr, the centres of mass of the stellar clumps along the filament (see Figure \ref{fig:hdbscan}) are separated by 2.0pc on average. We can compare the separation of these cores to that expected of an isothermal cylindrical gas distribution from the derivation presented in \citet{nagasawa}, namely:

\begin{equation}
\label{eq:fil}
    \lambda = 7.8 c_s \sqrt{\frac{2}{\pi G \rho_0}}
\end{equation}
where $\lambda$ is the separation between dense cores in a filament, c$_\mathrm{s}$ is the sound speed of the gas, G is the gravitational constant, and $\rho_0$ is the central gas density of the filament. Using the density and temperature thresholds for star formation to provide an upper limit (see Section \ref{sec:numerical_methods}) we calculate $\lambda$=5.1pc, a factor of $\approx$2.5 larger than the separation of the HDBSCAN clumps at t=1Myr. This calculation, however, does not take into account the changing density of the filament, or the external pressure from the surrounding environment. We discuss this more in Section \ref{sec:discussion}.

\begin{figure}
    \centering
    \includegraphics[width=0.95\linewidth, height=1\linewidth]{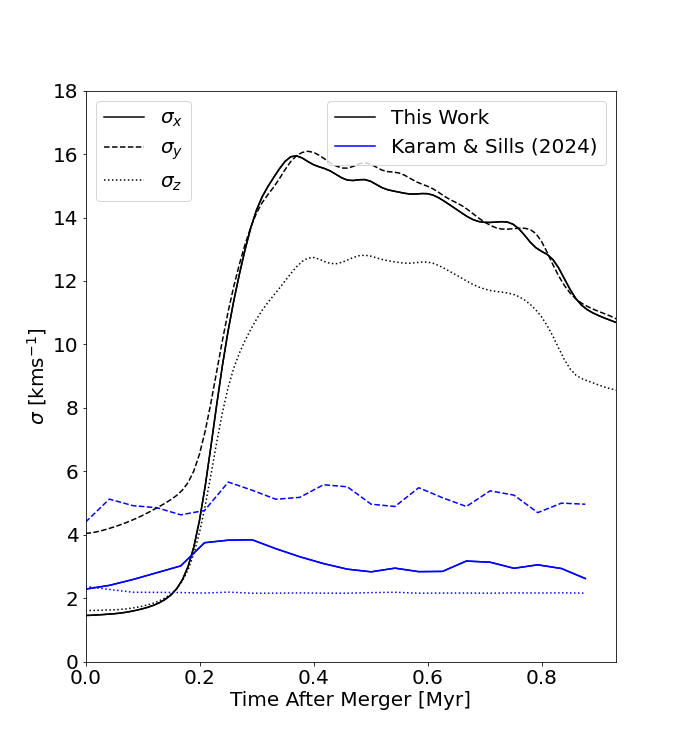}
    \caption{Velocity dispersion along each axis for cluster AB from the \texttt{region1} simulation in this paper (black) and in Paper I (blue).}
    \label{fig:sigma_AB}
\end{figure}

\begin{figure*}
    \centering
    \includegraphics[scale=0.35]{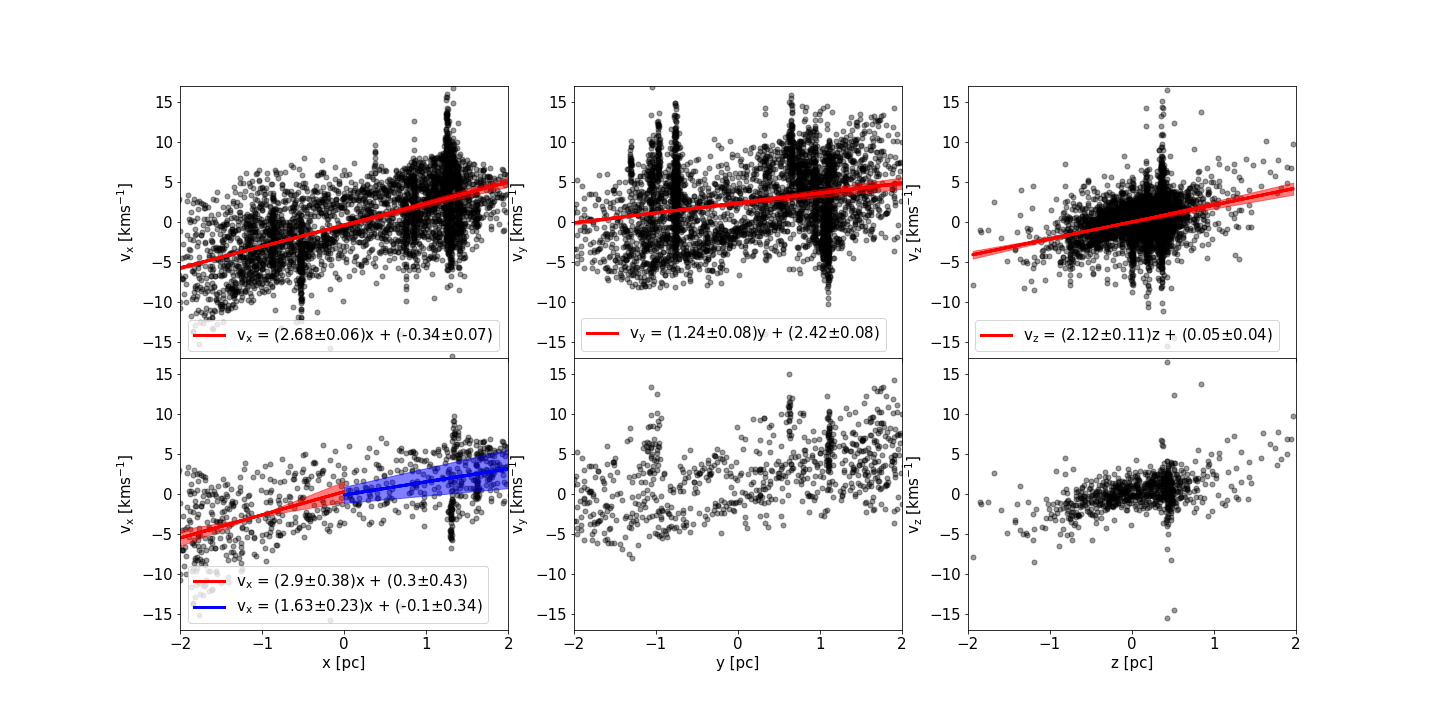}
    \caption{The position against velocity for stars belonging to cluster AB at distances between the 75\% mass radius and the nearest clump identified by HDBSCAN (top), and between the 90\% mass radius and the nearest clump identified by HDBSCAN (bottom). Left, middle, and right panels show this along the x, y, and z axes respectively. Red lines in the top panel show the lines of best fit data with the errors obtained through bootstrapping 10$^5$ times. Red and blue lines in the bottom panel show the lines of best fit for the data with x $<$ 0 and x $>$ 0 respectively. Shaded regions for all lines show the 3$\sigma$ uncertainty of the best fit line. Positions and velocities are around the centre of mass position and velocity respectively of cluster AB.}
    \label{fig:pv_AB}
\end{figure*}

\subsection{Stellar Dynamics}
\label{sec:dynamics_r1}
We see from Figure \ref{fig:hdbscan} that the structure of cluster AB in physical space is flattened along the z-axis. This pattern is also seen in the velocity structure of the cluster as can be seen in Figure \ref{fig:sigma_AB} where we find that the velocity dispersion along the z-axis is lower than that along the x and y axes. The trend present in the evolution of the velocity dispersion of cluster AB is an increase after the start of the merger, and a slow decrease throughout the remainder of the simulation. In Paper I, we used this to determine a violent relaxation timescale for the merger of cluster A and B of 0.4Myr. Because the velocity dispersion has not plateaued by the 1Myr mark shown in Figure \ref{fig:sigma_AB}, we conclude that star formation has extended the dynamical timescale of the merger process. As well, star formation has caused the velocity dispersion to increase compared to the \texttt{region1} simulation from Paper I. We find that the velocity dispersion of the gas that participates in the merger follows a similar trend to the stars in the \texttt{region1\_SFFB} simulation. This implies that the increase in velocity dispersion comes from the formation of stars with velocity inherited from the dense gas.

After the merger, the 75\% and 90\% mass radii increase while the 50\% mass radius remains constant implying that there is expansion present in the outer regions of the cluster (similar to Paper I). To analyze this expansion in detail, we show position-velocity diagrams of the cluster AB stars in the \texttt{region1\_SFFB} simulation at radii between the 75\% mass radius and the distance to the nearest clump identified by HDBSCAN in the top row of Figure \ref{fig:pv_AB}. We see that there are signals of expansion along each axis by fitting a line to the data and using the slope to tell us the rate of expansion. We find that the expansion rates shown in Figure \ref{fig:pv_AB} are higher than those calculated for the \texttt{region1} simulation in Paper I. As well, anisotropic expansion is present in Figure \ref{fig:pv_AB} with the strongest expansion signature along the x-axis and the weakest along the y-axis. Furthermore, there is anisotropic expansion in the outer regions of the cluster present along the x-axis. We see this when considering stars beyond the 90\% mass radius and within the distance to the nearest HDBSCAN clump as shown in the bottom row of Figure \ref{fig:pv_AB}. By fitting lines to the stars along the negative and positive x-axes separately, we see that the expansion rate along the negative x-axis is 1.80$\pm$0.34 times that along the positive x-axis. The anisotropic expansion comes from a combination of the rotation induced onto cluster AB from the surrounding gas, and the centre of mass motion of the cluster which is along the positive x-axis.

The vertical streaks present along each axis in the data shown in Figure \ref{fig:pv_AB} show the substructure of cluster AB. As the cluster is growing in mass after the merger, it accretes new clumps of stars that have formed along a filament (see Section \ref{sec:fils}) and these clumps do not coalesce with the rest of the stars that make up cluster AB at this point in the simulation and instead show up as vertical streaks in the position-velocity diagrams. We check whether the accreted clumps contribute to the overall substructure in cluster AB by calculating the Q parameter (\citealt{cart_Q}):
\begin{equation}
    Q = \frac{\Bar{m}}{\Bar{s}}
\end{equation}
where $\Bar{m}$ is the mean edge length of the branches of a minimum spanning tree created for the star cluster, and $\Bar{s}$ is the mean separation between stars divided by the cluster radius. A value of Q $<$ 0.8 implies substructure, while a value of Q $>$ 0.8 implies a smooth distribution with a radial density gradient. At t$=$1Myr, the Q parameter is 1.36 implying that cluster AB is heavily centrally concentrated and lacks detectable sub-structure. This is expected as the accreted clumps are low in size and mass compared to that of cluster AB.

\begin{figure}
    \centering
    \includegraphics[width=1\linewidth, height=0.95\linewidth]{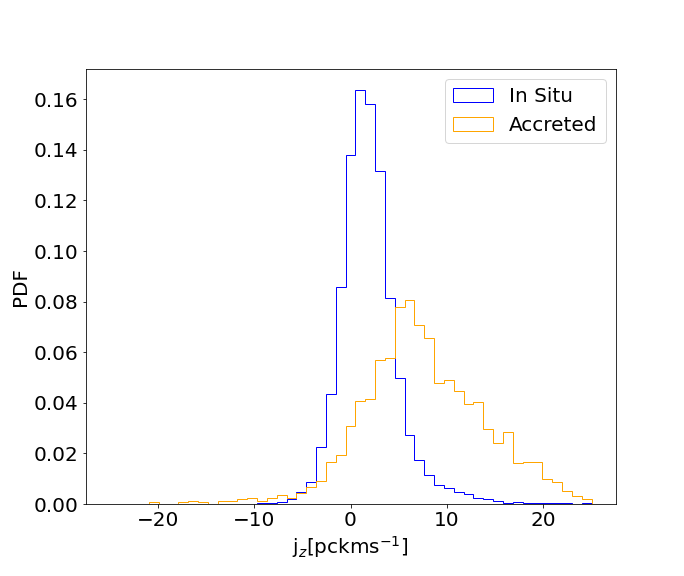}
    \caption{Distribution of z-component of specific angular momentum (j$_z$) for all stars belonging to cluster AB at t$=$1Myr in the \texttt{region\_1SFFB} simulation. Blue line shows stars which were formed in-situ of cluster AB, and orange line shows stars which were accreted onto cluster AB.}
    \label{fig:jz}
\end{figure}

We now analyze the change in angular momentum of cluster AB as the \texttt{region1\_SFFB} simulation evolves. We find that after the merger takes place, the total z-component specific angular momentum increases and continues doing so by t$=$1Myr after the start of the simulation. This is contrary to the results we found for the \texttt{region1} simulation in Paper I where the z-component specific angular momentum slowly decreased after reaching a peak at the beginning of the merger. To investigate the cause of this difference, we show the distribution of the z-component of the specific angular momentum for all cluster AB stars at t$=$1Myr in Figure \ref{fig:jz}. The blue histogram shows those stars which formed in-situ, and the orange histogram shows those stars which were accreted. We define an accreted star as a star which formed outside the 90\% mass radius of cluster AB, but was then bound to cluster AB at t$=$1Myr. 

From Figure \ref{fig:jz} we see that the distribution of the z-component specific angular momentum of the accreted stars is shifted compared to the same distribution for the in-situ stars by t$=$1Myr. The accreted stars account for 23\% of the total z-component specific angular momentum of cluster AB, but only account for 7\% of the total number of stars belonging to the cluster. This is a similar result to the \texttt{region2} simulation from Paper I and further illustrates the importance of the cluster environment in shaping the angular momentum of a merging star cluster. This result also shows that, while mergers can result in an increase in angular momentum, they are not the drivers of that increase in our simulations. Accretion of new stars from a filament or nearby cluster is necessary for the continuous supply of angular momentum in a cluster.

\subsection{Evolution Up To Gas Removal}

\begin{figure}
    \centering
    \includegraphics[width=1\linewidth, height=0.95\linewidth]{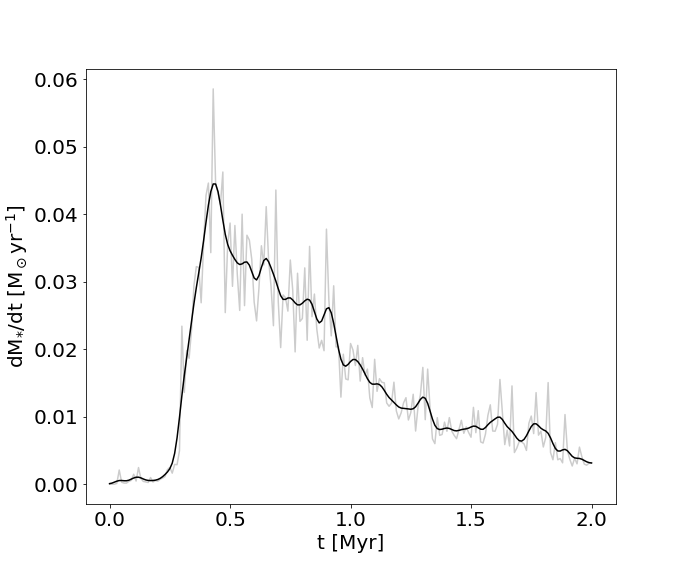}
    \caption{Star formation rate as a function of time throughout the \texttt{region1\_SFFB} simulation. Grey line shows original data and black line shows a Gaussian smoothing of the data over 0.02Myr.}
    \label{fig:SFR}
\end{figure}

By t$=$2Myr in the \texttt{region1\_SFFB} simulation, the bound stellar mass of cluster AB stops growing at 1.3$\times$10$^{4}$M$_\odot$. At this point, the mass of gas within cluster AB has decreased to 13\% of its original value. Because of this, we stop the simulation at t$=$2Myr. We also see that the star formation rate in the entire simulation has decreased drastically from its peak during the merger in Figure \ref{fig:SFR}. The crossing time of cluster AB at t$=$2Myr is 0.3Myr meaning that $\approx$3 crossing times have passed since t$=$1Myr. The half-mass relaxation time of cluster AB at 2Myr is 10.6Myr. The virial parameter ($\alpha = 2$K$/|\mathrm{U}|$ where K is the kinetic energy of the stellar component of the cluster and U is the potential energy of the stellar and gas component of the cluster) of cluster AB at this time is 1.4 implying that the cluster is supervirial. Though the cluster has survived only a very small fraction of its half-mass relaxation time, studying its dynamics at this phase of its evolution gives us crucial insight into the initial conditions for massive star cluster evolution after the clusters have emerged from their embedded phase.

We now discuss the survivability of the dynamical signatures we found to be inherited from the merger of cluster A with cluster B in the \texttt{region1\_SFFB} simulation (see Section \ref{sec:dynamics_r1}) at the time of gas removal in the cluster. At t$=$2Myr, cluster AB is still flattened along the z-axis in all of phase space. The velocity dispersions along the x and y axes remain $\approx$1.3 times higher than the velocity dispersion along the z-axis throughout the entire simulation after the onset of star formation triggered by the merger of cluster A and cluster B. We do not see any substructure in velocity space in cluster AB at t$=$2Myr.

By t$=$2Myr, the expansion rates in the outer regions of cluster AB along each axis have decreased from their maxima during the height of star formation and are 0.48$\pm$0.02Myr$^{-1}$, 0.36$\pm$0.02Myr$^{-1}$, 0.40$\pm$0.03Myr$^{-1}$ along x, y, and z respectively. The magnitude of the anisotropic expansion of cluster AB between x, y, and z present has decreased from its peak which occurred during the merger. We also calculate the strength of the anisotropic expansion along the x-axis and compare it to that present at t$=$1Myr (bottom left panel of Figure \ref{fig:pv_AB}). The expansion rate along the negative x-axis is 0.92$\pm$0.21Myr$^{-1}$ that along the positive x-axis which is lower than the same fraction calculated at t$=$1Myr. Therefore, as cluster AB has stopped growing through star formation and accretion, its rate of expansion has decreased and so too has the magnitude of the anisotropic expansion. However, both are still present in the cluster.

\begin{figure}
    \centering
    \includegraphics[width=1\linewidth, height=0.95\linewidth]{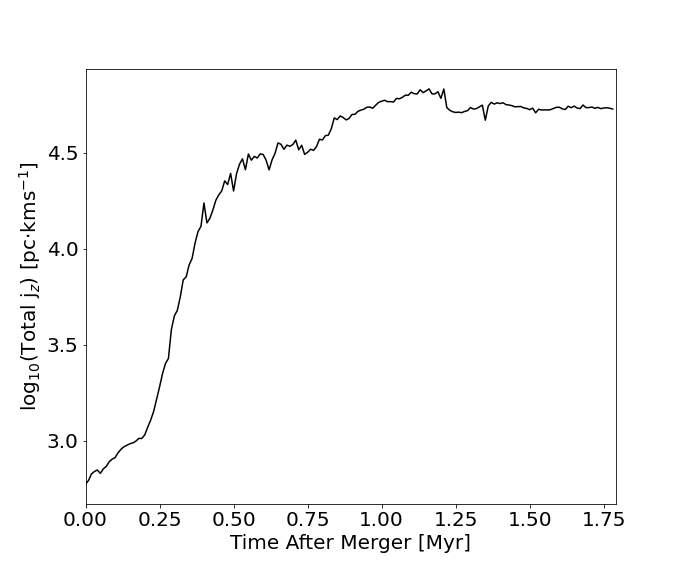}
    \caption{Total z-component specific angular momentum of stellar component of cluster AB in the \texttt{region1\_SFFB} simulation.}
    \label{fig:tot_jz}
\end{figure}

The z-component angular momentum of cluster AB decreases slightly from its peak value by 2Myr as seen in Figure \ref{fig:tot_jz}. This begins when clumps which were accreted onto cluster AB from the star forming filament split off of cluster AB instead of becoming fully accreted onto the cluster. The accreted stars still contribute a disproportionate amount to the z-component specific angular momentum though to a lesser extent than earlier on in the simulation. The accreted stars account for 6.5\% of the total number of stars belonging to cluster AB but account for 15\% of the total z-component specific angular momentum at t=2Myr. This is because the accreted stars remain in the outskirts of cluster AB after becoming accreted converse to the in-situ stars which remain closer to the cluster centre.

\begin{figure}
    \centering
    \includegraphics[width=1\linewidth, height=0.8\linewidth]{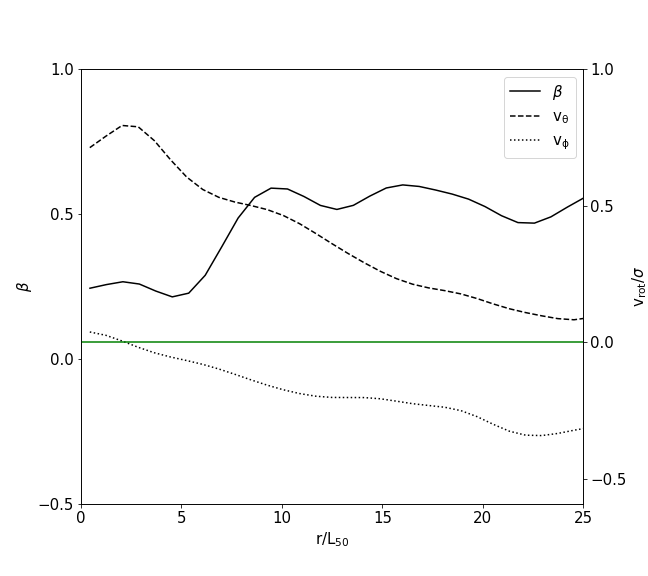}
    \caption{Velocity anisotropy (solid, left y-axis) and rotational velocity profiles of cluster AB in the \texttt{region1\_SFFB} simulation in spherical coordinates (dashed, and dotted, right y-axis) at t$=$2Myr. The rotational velocity is normalized by $\sigma = \sqrt{\sigma_r^2 + \sigma_\theta^2 + \sigma_\phi^2}/3$. All black lines show a Gaussian smoothing of the respective data over 0.5pc. The horizontal green line shows a value of zero.}
    \label{fig:dynamics}
\end{figure}

We also construct the rotational velocity profile of cluster AB by calculating the rotational velocity in spherical coordinates in evenly spaced bins from the centre of the cluster (Figure \ref{fig:dynamics}). The lack of rotation in the $\phi$ direction comes from the fact that the merger takes place mostly in the x-y plane resulting in counter clockwise rotation. The v$_\theta$ profile shows differential rotation is present inside the cluster with a profile that has also been seen in clusters that have experienced violent relaxation from core collapse: low rotation in the inner regions, a peak in rotational velocity in the intermediate regions, and lower to no rotational velocity in the outermost regions (e.g. \citealt{tiongco_rotation}). We discuss this more in Section \ref{sec:discussion}.

Lastly, to understand the future dynamical evolution of cluster AB, we calculate the velocity anisotropy defined as in \citet{binney_tremaine}:
\begin{equation}
    \beta = 1 - \frac{\sigma_{\phi}^2 + \sigma_{\theta}^2}{2\sigma_{r}^2}
\end{equation}
where $\sigma_{\phi}$ and $\sigma_{\theta}$ are the dispersions of the tangential component of the velocity, and $\sigma_r$ is the dispersion of the radial component of the velocity in spherical coordinates. The presence of velocity anisotropy has implications for the long term evolution of a cluster (e.g. \citealt{tiongco}, \citealt{pavlik_2021}, \citealt{pavlik_2024}) by controlling timescales upon which relaxation processes including core collapse take place. In the \texttt{region1\_SFFB} simulation we find that cluster AB begins developing very slight velocity anisotropy post merger and by 2Myr, has $\beta = $0.17 implying that it will begin its post embedded phase evolution with only mild radial anisotropy. We show the distribution of the velocity anisotropy as a function of radial distance away from the cluster centre in Figure \ref{fig:dynamics}. Qualitatively, this profile is similar to that presented in \citet{vesperini_2014}: an isotropic core, and a radially anisotropic intermediate region (note that by 2Myr, 10 half mass radii is equal to the 90\% mass radius). This trend is similar to that of the virial parameter which is much higher in the outer regions of the cluster than in the inner regions. The velocity anisotropy profile presented here does not return to isotropy in the outer regions like that from \citet{vesperini_2014} likely because our simulation does not include the long term evolution of the cluster inside an external tidal field (\citealt{baumgardt_makino}). In \citet{vesperini_2014}, this profile emerged as a response to the violent relaxation process felt by a cluster as it reaches core collapse alongside an external tidal field. As a star cluster merger can be described using violent relaxation (see Paper I), it is expected that we see a similar velocity anisotropy profile. Our results here show that evolution within the natal gas cloud is not enough to erase the signatures of violent relaxation in the outer regions of the cluster that were inherited from the cluster merger before the star cluster removes its surrounding gas.

\section{Discussion}
\label{sec:discussion}

\citet{FY21} show that stellar feedback is inefficient at suppressing star formation when cloud surface densities reach $\Sigma \approx$100M$_\odot$pc$^{-2}$. Such a suppression can lead to high star formation efficiencies $\epsilon_{\mathrm{*}}$= M$_\mathrm{*}$/M$_{\mathrm{gas,0}}$ where M$_{\mathrm{*}}$ is the stellar mass formed, and M$_{\mathrm{gas,0}}$ is the mass of gas at the beginning of the simulation. We calculate $\epsilon_{\mathrm{*}}$ at the end of our simulation and find $\epsilon_{\mathrm{*}}$=51\%. Such a high star formation efficiency is expected in our simulation as the gas distribution we use as our initial condition contains gas with $\Sigma \gtrapprox$1000M$_\odot$pc$^{-2}$ within each sub-cluster. Furthermore, as the merger of cluster A with cluster B increases gas density within the cluster, feedback is expected to be even less effective at stopping star formation.

The runaway stars produced in the \texttt{region1\_SFFB} simulation are almost all stars that were formed as a result of the compression of gas from the cluster merger. We do not find a preferred direction for the runaway stars converse to results found in simulations by \citet{Polak2024} and in observations by \citet{stoop_r136}. This is likely due to the fact that cluster A and cluster B in our simulation have similar masses, and in turn, similar potentials which is converse to the potential distribution from \citet{Polak2024}. As a consequence, we do not see the slingshot effect that was seen in \citet{Polak2024} to produce runaways with a preferred direction. Allowing for the formation of primordial binaries will also likely affect runaway star production (\citealt{CCC_JK}). 

%This lack of preferential direction of ejected stars is likely unique to mergers that take place in gas-rich environments (wet mergers) as they will result in high star formation in dense regions and, in turn, more random dynamical ejections from dynamically formed binaries. Conversely, dynamical ejections that result from cluster mergers in less gas-rich environments (dry mergers) may have a preferred direction because of the lack of new star formation. 

In Paper I, we predicted that a cluster may be able to develop kinematic subgroups associated with age gradients as they grow through mergers. We do not see these subgroups in the \texttt{region1\_SFFB} simulation presented in this work. The clumps that form along the nearby filament and become accreted onto cluster AB are too small compared to the amount of star formation that takes place within cluster AB during the merger and do not show up as distinct shapes in velocity space. However, if we only consider the HDBSCAN clumps that are not clusters AB or C (i.e. those that form along the filament), we do find the presence of subgroups in velocity space meaning that kinematic subgroups in clusters may preferentially arise in dry star cluster mergers.

The filament seen in Figure \ref{fig:vel_fil} is influenced by the surrounding gas within the cloud. As seen by the velocity streamlines in this figure, gas is flowing onto the filament from above and below its principal axis. This process acts as external pressure onto the filament which has a bearing on its core separation leading to smaller core separations than those expected from Equation \ref{eq:fil} (\citealt{nagasawa}). \citet{anathpindika} showed that high pressure environments can push core separations down to the order of 0.1pc. As well, the average gas density along the filament in our simulation presented here is greater than the star formation threshold density during star formation across the filament leading to a further decrease in the expected core separation. We therefore propose the following method of clump growth in filaments: the filament begins by forming stars at low separations, and these stars collect together to form the HDBSCAN identified clumps. In our scenario, gas cores form quickly after the formation of the filament, collapse to form stars, and the stars collect to form clumps. 

The addition of star formation and cooling into our simulation has reinforced a key conclusion from Paper I: the distribution of gas is extremely important in influencing the dynamics of clusters as they merge embedded inside a GMC. Our results from this work suggest that dynamical signatures that are inherited from the hierarchical cluster formation process are able to persist throughout the embedded phase of a clusters life and beyond.

We can use the rotation and velocity anisotropy present in cluster AB at t=2Myr in the \texttt{region1\_SFFB} simulation to predict its long-term evolution. We estimate the concentration c = $\log_{\mathrm{10}}(L_{\mathrm{90}}/L_{\mathrm{10}})$ where L$_\mathrm{90}$ is the 90\% mass radius, and L$_\mathrm{10}$ is the 10\% mass radius, of the cluster at t=2Myr and find that c$\approx$2. This along with the mass and size of cluster AB at t=2Myr suggests that it will likely be destroyed by evaporation in $\approx$40t$_\mathrm{rh}$ where t$_{\mathrm{rh}}$ is the half-mass relaxation time of cluster AB (\citealt{gnedin_ostriker}). Beginning with rotation, \citealt{livernois_2022} find that clusters with initial rotation can develop mass dependent rotation in as little as 1t$_{\mathrm{rh}}$ well before the expected evaporation of cluster AB. As the cluster evolves further, the distribution of angular momentum is expected to move outwards and the cluster will lose mass and memory of its initial rotation before its expected evaporation time (\citealt{tiongco_2017}). Lastly, the velocity anisotropy profile of cluster AB at t=2Myr suggests a quicker evolution towards core-collapse and energy equipartition than a cluster with an isotropic profile with both processes occurring well before the expected evaporation time (\citealt{pavlik_2021}).

The decrease in overall expansion present in cluster AB is consistent with observations presented in \citet{expansion_della_croce} where the authors observe a trend of decreasing expansion magnitude with increasing age. We cannot compare the expansion rates present in this work to those found in \citet{wright_2} because those clusters have much lower stellar mass than cluster AB in this work. Future work involving the simulation of star cluster formation from a wider array of initial gas morphologies will allow us to perform a comparison.

\section{Summary}
\label{sec:summary}
We have run a zoom-in simulation of star cluster assembly with the inclusion of star formation and a stellar feedback prescription to analyze the role played by star cluster mergers in shaping the dynamics of star clusters as they evolve embedded inside their natal gas cloud. We have compared this simulation to the same one without star formation or stellar feedback (\texttt{region1} from Paper I). Star cluster growth in the simulation with star formation and feedback happens via star formation within the merged cluster as a result of gas compression, and through accretion of nearby star forming gas from a filament. Clumps of stars have also formed along the filament and become accreted onto the merged cluster. In the simulation presented in this work with star formation and feedback, the magnitudes of the rotation and anisotropic expansion of the stellar component of the resultant cluster are stronger than in the same simulation from Paper I without star formation or feedback. Rotation is enhanced by gas accretion from a nearby filament, and this contributes to the anisotropic expansion towards the merger axis. We also analyze the dynamics of the resultant cluster after it has removed most of its surrounding gas through star formation and feedback. By the end of the simulation, the difference between the expansion rates along each axis has decreased implying the magnitude of the anisotropic expansion has decreased. Conversely, the angular momentum increases until t$\approx$1Myr and stays roughly constant until the end of the simulation with only slight decreases whenever clumps become removed from the merged cluster. The rotational velocity profile of the resultant cluster is peaked in the intermediate region of the cluster. We also analyze the velocity anisotropy profile of the cluster and find that the merged cluster has an isotropic core with a radially anisotropic outer region.

Because of the strong influence of gas morphology on cluster dynamics, we suggest that future simulations should employ an accurate modelling of the interstellar medium within which star formation and star cluster build up can occur if they are to accurately reproduce the emerging star cluster.

%% IMPORTANT! The old "\acknowledgment" command has be depreciated. It was
%% not robust enough to handle our new dual anonymous review requirements and
%% thus been replaced with the acknowledgment environment. If you try to 
%% compile with \acknowledgment you will get an error print to the screen
%% and in the compiled pdf.
%% 
%% Also note that the akcnowlodgment environment does not support long amounts of text. If you have a lot of people and institutions to acknowledge, do not use this command. Instead, create a new \section{Acknowledgments}.
\begin{acknowledgments}
The authors thank Takayuki Saitoh, Rachel Pillsworth, Claude Cournoyer-Cloutier, Veronika Dornan, and Chi Hong Lin for helpful discussions. The authors also thank the referee for careful reading and constructive comments. This project was the result of an award given to JK by the Japan Society for the Promotion of Science (JSPS) that was used to travel to Japan to work alongside MF. AS is supported by the Natural Sciences and Engineering Research Council of Canada. This research was enabled in part by support provided by Compute Ontario (https://www.computeontario.ca), Compute Canada (http://www.computecanada.ca), and the Center for Computational Astrophysics at the National Astronomical Observatory of Japan (https://www.cfca.nao.ac.jp/en).
\end{acknowledgments}

%% To help institutions obtain information on the effectiveness of their 
%% telescopes the AAS Journals has created a group of keywords for telescope 
%% facilities.
%
%% Following the acknowledgments section, use the following syntax and the
%% \facility{} or \facilities{} macros to list the keywords of facilities used 
%% in the research for the paper.  Each keyword is check against the master 
%% list during copy editing.  Individual instruments can be provided in 
%% parentheses, after the keyword, but they are not verified.

%% Similar to \facility{}, there is the optional \software command to allow 
%% authors a place to specify which programs were used during the creation of 
%% the manuscript. Authors should list each code and include either a
%% citation or url to the code inside ()s when available.

\software{astropy \citep{2013A&A...558A..33A,2018AJ....156..123A},  
          numpy \citep{numpy},
          matplotlib \citep{matplotlib},
          pynbody \citep{pynbody},
          Cloudy \citep{2013RMxAA..49..137F}, 
          scipy \citep{2020SciPy-NMeth}
          }

%% Appendix material should be preceded with a single \appendix command.
%% There should be a \section command for each appendix. Mark appendix
%% subsections with the same markup you use in the main body of the paper.

%% Each Appendix (indicated with \section) will be lettered A, B, C, etc.
%% The equation counter will reset when it encounters the \appendix
%% command and will number appendix equations (A1), (A2), etc. The
%% Figure and Table counter will not reset.

%% For this sample we use BibTeX plus aasjournals.bst to generate the
%% the bibliography. The sample631.bib file was populated from ADS. To
%% get the citations to show in the compiled file do the following:
%%
%% pdflatex sample631.tex
%% bibtext sample631
%% pdflatex sample631.tex
%% pdflatex sample631.tex

\bibliography{sample631}{}
\bibliographystyle{aasjournal}

%% This command is needed to show the entire author+affiliation list when
%% the collaboration and author truncation commands are used.  It has to
%% go at the end of the manuscript.
%\allauthors

%% Include this line if you are using the \added, \replaced, \deleted
%% commands to see a summary list of all changes at the end of the article.
%\listofchanges

\end{document}